\documentclass[aps,prc,twocolumn,superscriptaddress,showpacs]{revtex4}

\newcommand{\rpg}{($p$,$\gamma$)}
\newcommand{\rpp}{($p$,$p$)}
\newcommand{\rdp}{($d$,$p$)}
\newcommand{\rdn}{($d$,$n$)}
\newcommand{\rggp}{($\gamma$,$\gamma'$)}

\begin{document}

\title{
  Addendum to ``Determination of $\gamma$-ray widths in $^{15}$N using nuclear resonance fluorescence''
}

\author{Tam\'{a}s Sz\"{u}cs}
\email{t.szuecs@hzdr.de}
\affiliation{%
Helmholtz-Zentrum Dresden - Rossendorf (HZDR), D-01328 Dresden, Germany}
\author{Peter Mohr}
\affiliation{%
Diakonie-Klinikum, D-74523 Schw\"abisch Hall, Germany}
\affiliation{%
Institute for Nuclear Research (MTA Atomki), H-4001 Debrecen, Hungary}

\date{August 18, 2015}

\begin{abstract}
The determination of absolute widths of two observed levels above the proton threshold in $^{15}$N has been improved by a combined analysis of our recent $^{15}$N\rggp $^{15}$N$^\ast$ photon scattering data, resonance strengths 
$\omega\gamma$ of the $^{14}$C\rpg $^{15}$N reaction, and $\gamma$-ray branchings $b_{\gamma,i}$ in $^{15}$N. The revised data are compared to the adopted values, and some inconsistencies in the adopted values are illustrated.
\end{abstract}


\maketitle
In a recent study \cite{Szu15} photon scattering was used to determine level properties in $^{15}$N. A clear signal was observed from two levels above the adopted proton threshold in $^{15}$N at $S_p = 10207.4$\,keV. In the following all adopted values are taken from the ENSDF online database \cite{ENSDF} which is in general based on the TUNL update \cite{TUNL} of the latest compilation by Ajzenberg-Selove in 1991 \cite{Ajz91}. Separation energies are in agreement with the latest AME atomic mass evaluation \cite{Wang12}. \vspace{-.4mm}

The present analysis focuses on two $J = 3/2$ levels at $E_x = 10702$\,keV and $10804$\,keV in $^{15}$N. The partial radiation width to the ground state $\Gamma_{\gamma,0}$, the total radiation width $\Gamma_\gamma$, the proton width $\Gamma_p$, and the total width $\Gamma$ were derived from the experimental photon scattering data in our recent paper \cite{Szu15} in combination with resonance strengths $\omega\gamma$ in the $^{14}$C\rpg $^{15}$N reaction and $\gamma$-ray branchings $b_{\gamma,i}$ in $^{15}$N. In our previous analysis the small proton partial width of these states was not properly taken into account. The improved re-analysis of the present study leads to some interesting discrepancies with the adopted values which have been considered as certain over the last decades.  \vspace{-.4mm}

This study is organized as follows. As the determination of absolute widths in $^{15}$N is based on the combination of our recent photon scattering data \cite{Szu15} and data from literature, in the first part the required literature data are re-evaluated. In particular, these required data are the resonance strengths $\omega \gamma$ of the $^{14}$C\rpg $^{15}$N reaction and the $\gamma$-ray branchings $b_{\gamma,i}$ in $^{15}$N. The second and main part of the paper describes the determination of absolute widths $\Gamma_{\gamma,0}$, $\Gamma_\gamma$, $\Gamma_p$, and $\Gamma$ from the combination of our new photon scattering data and the re-evaluated literature data. Special attention is paid to the error propagation. In the third and
last part of this study the new results are compared to other data from literature, and surprisingly discrepancies to the adopted values are found in some cases. These discrepancies will be discussed in further detail. The new results are summarized in Table \ref{tab:level}.
\begin{table*}[tbh]
\caption{\label{tab:level}
Absolute widths and further properties of the two $J = 3/2$ levels in $^{15}$N at $E_x = 10702$\,keV and 10804\,keV. Note that all calculated results are given with a precision of at least 4 digits to avoid rounding errors; the number of significant digits is smaller (typically 2) and can be seen from the given uncertainties.
}
\begin{ruledtabular}\
\begin{tabular}{*{11}{c}}
$E_x$	& $J_x$	& $A = \Gamma_{\gamma,0}^2/\Gamma$ & $\omega \gamma$	& $\omega \gamma_0$	& $R_0$	& $b_{\gamma_0}$		& $\Gamma_{\gamma_0}$		& $\Gamma_{\gamma}$	& $\Gamma_{p}$	& $\Gamma$ \\
(keV)	&$-$		& (meV)					& (meV)			& (meV)			& ($-$)	& (\%)			& (meV)				& (meV)			& (meV)		& (meV) \\
\multicolumn{2}{c}{Refs.~\cite{ENSDF,TUNL,Ajz91}} & Ref.~\cite{Szu15} & \multicolumn{2}{c}{GOE-NPA}	& \multicolumn{6}{c}{this work: combination of various data; further details see text} \\
\colrule
10702 & $3/2$	& 215.8$\pm$17.2 				& 840$\pm$130 		& 352.8$\pm$64.1 		& 1.223$\pm$0.243 & 50.5$\pm$1.7 	& 603.7$\pm$48.9 			& 1195.4$\pm$101.6 	& 493.6$\pm$75.3	& 1689.0$\pm$126.4 \\
10804 & $3/2$ 	& 103.8$\pm$11.4				& 270$\pm$40\phantom{0}& 118.8$\pm$20.6		& 1.747$\pm$0.360 & 49.2$\pm$1.7	& 270.4$\pm$26.5			& 549.6$\pm$56.0		& 154.8$\pm$23.3	&   704.4$\pm$60.6  \\
\end{tabular}
\end{ruledtabular}
\end{table*}

Let us now start with the re-evaluation of the literature data for the resonance strengths $\omega \gamma$ and the $\gamma$-ray branchings $b_{\gamma,i}$.
\noindent 
Resonance strengths $\omega \gamma$: \\ 
The resonance strengths $\omega \gamma = \omega \Gamma_p \Gamma_\gamma / \Gamma$ of the $^{14}$C\rpg $^{15}$N reaction have been adopted in \cite{ENSDF,TUNL,Ajz91} from G\"orres {\it et al.}\ \cite{Goe90} (hereafter: GOE-NPA). Earlier measurements for the two levels under study have been done by Hebbard and Dunbar \cite{Heb59} (HD-PR), Siefken {\it et al.}\ \cite{Sie69} (SIE-NPA), and Beukens \cite{Beu76} (BEU-PhD); only a minor part of the Ph.D.\ thesis BEU-PhD has been published \cite{Beu75}. As there is no good agreement between the data by HD-PR, SIE-NPA, and BEU-PhD, and the unpublished data of BEU-PhD have been normalized to one particular resonance strength of
SIE-NPA, we confirm to adopt the latest resonance strengths $\omega \gamma$ by GOE-NPA: $\omega \gamma (10702) = 840 \pm 130$\,meV and $\omega \gamma (10804) = 270 \pm 40$\,meV. Note that $\omega = 2$ for the two $J = 3/2$ resonances.

\noindent 
Ground state resonance strengths $\omega \gamma_0$: \\
The partial resonance strengths $\omega \gamma_0 = \omega \Gamma_p \Gamma_{\gamma,0} / \Gamma$ of $^{14}$C\rpg $^{15}$N are also taken from GOE-NPA. The experimental quantities in GOE-NPA are the $\gamma$-ray yields for the transitions to the $i$-th excited state in $^{15}$N. Therefore it is consistent to use here the given resonance strengths $\omega \gamma$ and the given ground state $\gamma$-ray branchings $b_{\gamma,0}$ of the same experimental work. This leads to $\omega \gamma_0(10702) = 352.8 \pm 64.1$\,meV from $b_{\gamma,0} = 0.42 \pm 0.04$ for the 10702\,keV state and $\omega \gamma_0 = 118.8 \pm 20.6$\,meV from $b_{\gamma,0} = 0.44 \pm 0.04$ for the 10804\,keV state. 
The uncertainties of the values come from the uncertainty of the resonance strengths and the uncertainty of the branching ratio from the same experiment. This overestimates the uncertainty of the actually measured $\omega \gamma_0$, because in the original work the branching ratios were derived from the measured partial strengths. But without further information on the error estimate for partial resonance strengths in GOE-NPA this choice seems to be a careful compromise.
The ratio $\Gamma_p \Gamma_{\gamma,0}/\Gamma$ is a factor of $\omega = 2$ smaller than the above quoted numbers for $\omega \gamma_0$.

\noindent 
Ground state $\gamma$-ray branches $b_{\gamma,0}$: \\ 
$\gamma$-ray branchings have been adopted in \cite{ENSDF,TUNL,Ajz91} from the unpublished BEU-PhD data because of their very small uncertainties. However, the BEU-PhD data and the later GOE-NPA data have both been derived from 
$^{14}$C\rpg $^{15}$N experiments, and the later GOE-NPA data have smaller uncertainties for the resonance strengths, but larger uncertainties for the branching ratios; this leaves some doubt on the very small uncertainties of BEU-PhD. A weighted average of the three experiments with high-resolution detectors (SIE-NPA, BEU-PhD, GOE-NPA) is dominated by the tiny uncertainties of BEU-PhD and leads to $b_{\gamma,0}(10702) = (51.12 \pm 0.62_{\rm{int}} \pm 1.14_{\rm{ext}})$\,\% and $b_{\gamma,0}(10804) = (50.82 \pm 0.37_{\rm{int}} \pm 1.18_{\rm{ext}})$\,\%. The unweighted average gives significantly lower values of $b_{\gamma,0}(10702) = 48.87$\,\% and $b_{\gamma,0}(10804) = 47.50$\,\%. We finally adopt the average of the above numbers with an estimated $1\sigma$ uncertainty which includes the higher weighted average and the lower unweighted average: $b_{\gamma,0}(10702) = (50.5 \pm 1.7)$\,\% and $b_{\gamma,0}(10804) = (49.2 \pm 1.7)$\,\%.
Note if we use the adopted $\gamma$-ray branchings throughout the analysis, the final results will be reduced by about $3-8$\,\%, but will still remain within the given error bars.
\\

Now the required data from literature are fixed, and we can combine the above literature data with our new data for the integrated photon scattering cross sections $I_\sigma$ from the $^{15}$N\rggp $^{15}$N$^\ast$ experiment. This will allow
to fix all widths. The integrated cross section for the transition $0 \rightarrow J_x,E_x \rightarrow 0$ in $^{15}$N\rggp $^{15}$N$^\ast$ photon scattering is given by:
\begin{equation}
I_\sigma(0 \rightarrow J_x,E_x \rightarrow 0) = 
\frac{2J_x + 1}{2J_0 + 1} \left( \frac{\pi \hbar c}{E_x} \right)^2
\frac{\Gamma_{\gamma,0} \Gamma_{\gamma,0}}{\Gamma}
\label{eq:intsig}
\end{equation}
with the ground state $\gamma$-ray branching $b_{\gamma,0} = \Gamma_{\gamma,0}/\Gamma_\gamma$ and the total width $\Gamma = \Gamma_p + \Gamma_\gamma$ for the states under consideration above the proton threshold and below the neutron threshold. Therefore, the value of $\Gamma_{\gamma,0}^2/\Gamma$ is fixed from our photon scattering data \cite{Szu15}.

\noindent 
The ratio $R_0 = \Gamma_{\gamma,0}/\Gamma_p$: \\
The integrated photon scattering cross section $I_\sigma$ for the $0 \rightarrow J_x,E_x \rightarrow 0$ transition is proportional to the quantity $\Gamma_{\gamma,0}^2/\Gamma$  (hereafter $A$), and the ground state resonance strength $\omega \gamma_0$ is proportional to $\Gamma_{\gamma,0} \Gamma_p/\Gamma$  (hereafter $B$). Thus, the ratio $R_0 = \Gamma_{\gamma,0}/\Gamma_p = A/B$ can directly be derived from the ratio of the above two quantities $I_\sigma$ from \cite{Szu15} and $\omega \gamma_0$ from GOE-NPA. The results $R_0(10702) = 1.223 \pm 0.243$ and $R_0(10804) = 1.747 \pm 0.360$ clearly show that the proton width $\Gamma_p$ is smaller than the radiation width $\Gamma_\gamma$ for both levels under study. This result is independent of the spin assignment $J$ of the levels.

\noindent 
The partial radiation width to the ground state $\Gamma_{\gamma,0}$: \\
The quantity $\Gamma_{\gamma,0}^2/\Gamma = \Gamma_{\gamma,0}^2/(\Gamma_p + \Gamma_\gamma)$ from the integrated photon scattering cross section $I_\sigma$ can be combined with the ground state $\gamma$-ray branching $b_{\gamma,0} = \Gamma_{\gamma,0}/\Gamma_\gamma$ and the above ratio $R_0 = \Gamma_{\gamma,0}/\Gamma_p$. This leads to 
\begin{equation}
\Gamma_{\gamma,0} = 
\left( \frac{\Gamma_{\gamma,0}^2}{\Gamma} \right) \,
\left( \frac{R_0 + b_{\gamma,0}}{R_0 \times b_{\gamma,0}} \right) =
A \,
\left( \frac{1}{b_{\gamma,0}} + \frac{1}{R_0} \right)
\label{eq:gam0}
\end{equation}
where $A$ is taken from $I_\sigma$ from the photon scattering data \cite{Szu15}, and the numbers $R_0$ and $b_{\gamma,0}$ in the parenthesis have been determined above.
To avoid double counting the uncertainties of the widths, the shown equations are transformed to be dependent only on the independent experimental values with known uncertainties and not on the correlating derived values, i.\,e.\ Eq.~(\ref{eq:gam0}) becomes
\begin{equation}
\Gamma_{\gamma,0} = 
A \, \frac{1}{b_{\gamma,0}} +  B
\label{eq:gam0_mod}
\end{equation}
where $A$ is taken from $I_\sigma$ from the photon scattering data \cite{Szu15}, $B$ is from $\omega \gamma_0$ from GOE-NPA and $b_{\gamma,0}$ have been determined above. This leads to $\Gamma_{\gamma,0}(10702) = 603.7 \pm 48.9$\,meV and $\Gamma_{\gamma,0}(10804) = 270.4 \pm 26.5$\,meV. 

\noindent 
The total radiation width $\Gamma_\gamma$: \\
The total radiation width $\Gamma_\gamma$ is directly related to the ground state radiation width $\Gamma_{\gamma,0}$ by $b_{\gamma,0} = \Gamma_{\gamma,0}/\Gamma_\gamma$. From the above numbers we find $\Gamma_\gamma(10702) = 1195.4 \pm 101.6$\,meV and $\Gamma_\gamma(10804) = 549.6 \pm 56.0$\,meV.

\noindent 
The proton width $\Gamma_p$: \\
The proton width $\Gamma_p$ is directly related to the ground state radiation width $\Gamma_{\gamma,0}$ by $R_0 = \Gamma_{\gamma,0}/\Gamma_p$. From the above numbers we find $\Gamma_p(10702) = 493.6 \pm 75.3$\,meV and $\Gamma_p(10804) = 154.8 \pm 23.3$\,meV.

\noindent 
The total width $\Gamma$: \\
Finally, the total width $\Gamma$ can simply be calculated by the sum of the partial widths: $\Gamma = \Gamma_p + \Gamma_\gamma$. The obtained values are $\Gamma(10702) = 1689.0 \pm 126.4$\,meV and $\Gamma(10804) = 704.4 \pm 60.6$\,meV. The relative uncertainties of the total widths $\Gamma$ ($\approx 7$ and $9$\,\%) remain smaller than for $\Gamma_p$ because of $\Gamma_\gamma > \Gamma_p$ and the smaller uncertainties of $\Gamma_\gamma$.
The given uncertainties were calculated using standard error propagation. However, this may slightly underestimate the real uncertainties of $\Gamma$ because $\Gamma_\gamma$ and $\Gamma_p$ are not statistically independent. A more realistic estimate is about 10\,\%, i.e.\ similar to the uncertainty of $\Gamma_\gamma$.
\\

Next, we compare the absolute widths from the present study to the adopted values \cite{ENSDF,TUNL,Ajz91}. In addition, we try to trace back to the
origins of the adopted values.

\noindent
The 10804\,keV state: \\ 
In the ``Energy levels of $^{15}$N'' table 15.4 of Ajzenberg-Selove 1991 (hereafter Ajz91; \cite{Ajz91}) one finds $J^\pi = 3/2^+$ and $\Gamma < 1$\,eV. In Table 15.11 ``Resonances in $^{14}$C + $p$'' of Ajz91 one finds $J^\pi = 3/2^{(+)}$, and in addition $\Gamma_p = 220 \pm 100$\,meV, and $\Gamma_\gamma = 270 \pm 140$\,meV with a footnote $\omega \gamma = 270 \pm 40$\,meV (GOE-NPA). This is a minor inconsistency in Ajz91 because the combination of $\omega \gamma = 270$\,meV from GOE-NPA and the adopted $\Gamma_p = 220$\,meV leads to $\Gamma_\gamma = 350$\,meV; the adopted lower value of $\Gamma_\gamma = 270$\,meV in Ajz91 is only obtained if the earlier resonance strength $\omega \gamma = 240$\,meV from BEU-PhD is used. The values $\Gamma_p = 220 \pm 100$\,meV and $\Gamma_\gamma = 270$\,meV are also found in earlier versions of Ajzenberg-Selove (Ajz86 \cite{Ajz86}, Ajz81 \cite{Ajz81}, Ajz76 \cite{Ajz76}), and in Ajz76 BEU-PhD is explicitly given as reference. The adopted $\Gamma_p = 220 \pm 100$\,meV is derived in BEU-PhD from the measured resonance strength $\omega \gamma$ and the ratio $\Gamma_\gamma/\Gamma = 0.55^{+0.25}_{-0.15}$ from a detailed study of electromagnetic transitions in $A = 15$ nuclei by Warburton {\it et al.}\ \cite{War65}. Earlier compilations (Ajz70 \cite{Ajz70} and Ajz59 \cite{Ajz59}) give only the resonance strength from earlier work, but no partial widths $\Gamma_\gamma$ or $\Gamma_p$. 

The proton width $\Gamma_p = 154.8 \pm 23.3$\,meV of this study compares well to the adopted value of $220 \pm 100$\,meV but has a significantly reduced uncertainty. This allows to determine the total width $\Gamma = 704.4 \pm 60.6$\,meV which is consistent with the previous upper limit of 1\,eV. The radiation width $\Gamma_\gamma$ of this study is about a factor of two higher than the adopted value. The present results are also consistent with the result of Warburton {\it et al.}\ \cite{War65}: The present study finds $\Gamma_\gamma/\Gamma = 0.78$ and $\Gamma_{\gamma,0}/\Gamma = 0.38$, in agreement with the respective values
of $0.55^{+0.25}_{-0.15}$ and $0.30^{+0.15}_{-0.09}$ of \cite{War65}.

The 10804\,keV state has not been seen in proton transfer in the $^{14}$C\rdn $^{15}$N reaction \cite{Bom75} or in $^{14}$C\rpp $^{14}$C elastic scattering \cite{Heb59}. This is again consistent with a small proton width $\Gamma_p$.

For completeness it has to be noted that instead of the adopted $J^\pi = 3/2^+$ in Ajz91 $J^\pi = 3/2^-$ is reported earlier in the ``Energy levels'' table of Ajz59 which is based on $\gamma$-ray angular distribution measurements in \cite{Bar54,Bar55}. The experimental data of \cite{Bar55} clearly show that $J = 3/2$ and prefer $J^\pi = 3/2^-$ but cannot exclude $J^\pi = 3/2^+$. The value $J^\pi = 3/2^-$ from Ajz59 changes to $3/2^{(-)}$ in Ajz70, $3/2^{(+)}$ in Ajz76 (probably again based on BEU-PhD), and $3/2^+$ in Ajz81, Ajz86, Ajz91. However, $J^\pi = 3/2^{(+)}$ persists in the ``Resonances in $^{14}$C + $p$'' tables up to Ajz91. The analysis of the angular distribution of the $^{14}$N\rdp $^{15}$N reaction in \cite{Phi69} shows clear signature of $L=1$ transfer, i.\,e. it indicates negative parity of this state. As the spin $J = 3/2$ is well-defined from $\gamma$-ray angular distribution and angular correlation measurements by Bartholomew {\it et al.}\ \cite{Bar55} and SIE-NPA, $J^\pi = 3/2^-$ should be adopted instead of $J^\pi = 3/2^+$. Surprisingly, an electron scattering experiment reports conflicting results with $J^\pi = 3/2^+$ and $\Gamma^{\rm{M2}}_{\gamma,0} = 18 \pm 8$\,meV (see Table 15.17 in Ajz81, based on the experimental data of \cite{Ans77}).

\noindent
The 10702\,keV state: \\
The situation for the 10702\,keV state is even worse than for the 10804\,keV state. Ajz91 gives $J^\pi = 3/2^-$ and $\Gamma = 0.2$\,keV in the ``Energy levels of $^{15}$N'' table; the same numbers are found in Ajz86, Ajz81, and Ajz76. Ajz70 and Ajz59  state $J^\pi = 3/2^+$, based on angular correlation measurements in \cite{Bar54,Bar55} and $^{14}$C\rpp $^{14}$C elastic scattering data from \cite{Heb59}. Similar to the 10804\,keV state, the 1976 change of the adopted values is based on BEU-PhD, and the ``Resonances in $^{14}$C + $p$'' table provides in addition $\Gamma = 0.2$\,keV and $\Gamma_\gamma = 370 \pm 70$\,meV; the latter value is taken from $\omega \gamma = 740 \pm 140$\,meV in BEU-PhD and $\Gamma_\gamma \ll \Gamma_p \approx \Gamma$ and is kept until Ajz91 (again, a footnote in Ajz91 states $\omega \gamma = 840 \pm 130$\,meV from GOE-NPA, but this value is not used in Ajz91 for further calculations).

The huge adopted proton width of $\Gamma_p = 0.2$\,keV is a factor of about 400 above the result of the present study ($\Gamma_p = 493.6 \pm 75.3$\,meV). A state with such a huge proton width would not be visible in photon scattering because such a state decays preferentially by proton emission because of $\Gamma_p \gg \Gamma_{\gamma,0}$. Thus, $\Gamma_p = 0.2$\,keV for the 10702\,keV state is clearly ruled out by the present study.

A claim for the huge proton width of $\Gamma_p = 0.2$\,keV has been made from the $^{14}$C\rpp $^{14}$C elastic scattering data of \cite{Heb59}. The proton width was estimated from the deviation of the elastic scattering cross section from Rutherford scattering, and these data were also used to pin down the positive parity of the 10702\,keV state, leading to the adopted $J^\pi = 3/2^+$ in Ajz59 and Ajz70. As estimated in \cite{Heb59}, the large proton width corresponds to about 20\,\% of the Wigner limit. Such a strong state should be clearly visible in the $^{14}$C\rdn $^{15}$N transfer experiment \cite{Bom75}, but also the 10702\,keV state has not been detected in \cite{Bom75}. Therefore, the claim for the huge proton width $\Gamma_p = 0.2$\,keV and the positive parity of the 10702\,keV state from \cite{Heb59} seems to be not well-founded. Nevertheless, we finally recommend to adopt $J^\pi = 3/2^+$ because the $^{14}$N\rdp $^{15}$N data of \cite{Phi69} show
clear signature of $L = 2$ transfer with a small contribution of $L = 0$ transfer, i.e.\ clear signature of a positive parity of the 10702\,keV state. Further confirmation of the positive parity of the 10702\,keV state is taken from the analysis of thermal neutron capture of $^{14}$N by Jurney {\it et al.}\ \cite{Jur97}.

The condition $\Gamma_p \gg \Gamma_\gamma$ does not hold for the newly derived $\Gamma_p$ from this study. Consequently, earlier adopted values for $\Gamma_\gamma$ are also inconsistent because they were derived from the resonance strength using $\omega \gamma \approx \omega \Gamma_\gamma$ which is not a valid approximation in the present case.

In conclusion, the present study has determined absolute widths $\Gamma_{\gamma,0}$, $\Gamma_\gamma$, $\Gamma_p$, and $\Gamma$ for the two $J = 3/2$ states in $^{15}$N at $E_x = 10702$\,keV and 10804\,keV from a
combination of integrated cross sections $I_\sigma$ from $^{15}$N\rggp $^{15}$N$^\ast$ and from resonance strengths $\omega \gamma$ and $\gamma$-ray branchings $b_{\gamma,i}$ from $^{14}C$\rpg $^{15}$N. For the 10804\,keV state the results are roughly consistent with the adopted values \cite{ENSDF,TUNL,Ajz91} but have significantly lower uncertainties. Contrary, the proton width $\Gamma_p$ of the 10702\,keV state is about a factor of 400 lower than the adopted value; this affects also the earlier estimates of the radiation widths of this state which are based on the incorrect assumption $\Gamma_p \gg \Gamma_\gamma$. Furthermore, the parity assignments of both states should be changed to $J^\pi = 3/2^+$ for the 10702\,keV state and $J^\pi = 3/2^-$ for the 10804\,keV state.

\noindent
%
This work was supported by the Helmholtz Association through the Nuclear Astrophysics Virtual Institute (NAVI, Grant No. HGF VH-VI-417) and OTKA (Grant No. K101328 and No. K108459). 
%

\end{document}